\documentclass[aps,pre,twocolumn,groupedaddress]{revtex4-2}
\usepackage{amsmath,amssymb}	         
\usepackage{bm}				 
\usepackage{color}                       
\usepackage{graphicx}                    
\usepackage{algorithm,algpseudocode}     

\begin{document}

\newcommand{\myxLL}{\bm{x}} 
\newcommand{\myxL}{\bm{\xi}} 
\newcommand{\myg}{\bm{g}} 
\newcommand{\myz}{\bm{z}} 
\newcommand{\piL}{\pi_{\xi}} 
\newcommand{\tpiLL}{\tilde{\pi}_{x}} 
\newcommand{\piLL}{\pi_{x}} 


\title{Super-resolution of spin configurations based on flow-based generative models}


\author{Kenta Shiina}
\altaffiliation[Also at ]{Bioinformatics Institute, Agency for Science, Technology and Research (A*STAR), 30 Biopolis Street, No. 07-01 Matrix, 138671, Singapore.}
\email[e-mail: ]{16879316kenta@gmail.com}
\author{Hiroyuki Mori}
\author{Yutaka Okabe}
\affiliation{Department of Physics, Tokyo Metropolitan University, Hachioji, Tokyo 192-0397, Japan.}
\author{Lee Hwee Kuan}
\affiliation{Bioinformatics Institute, 138671, Singapore,}
\affiliation{School of Computing, National University of Singapore, 117417, Singapore,}
\affiliation{Singapore Eye Research Institute (SERI), 168751, Singapore,}
\affiliation{Image and Pervasive Access Laboratory (IPAL), 138632, Singapore,}
\affiliation{Rehabilitation Research Institute of Singapore, 308232, Singapore.}
\affiliation{Singapore Institute for Clinical Sciences (SICS), 117609, Singapore,}
\author{Yusuke Tomita}
\affiliation{College of Engineering, Shibaura Institute of Technology, Saitama 330-8570, Japan}


\date{\today}

\begin{abstract}
 We present a super-resolution method for spin systems using a flow-based generative model that is a deep generative model with reversible neural network architecture.
 Starting from spin configurations on a two-dimensional square lattice, our model generates spin configurations of a larger lattice.
 As a flow-based generative model precisely estimates the distribution of the generated configurations,
 it can be combined with Monte Carlo simulation to generate large lattice configurations according to the Boltzmann distribution.
 Hence, the long-range correlation on a large configuration is reduced into the shorter one through the flow-based generative model.
 This alleviates the critical slowing down near the critical temperature.
 We demonstrated an $8$ times increased lattice size in the linear dimensions using our super-resolution scheme repeatedly.
 We numerically show that by performing simulations for $16\times 16$ configurations, our model can sample lattice configurations at $128\times 128$
 on which the thermal average of physical quantities has good agreement with the one evaluated by the traditional Metropolis-Hasting Monte Carlo simulation.
\end{abstract}


\maketitle

 \section{Introduction}
 \label{sec:intro}
 In the field of condensed matter physics,
 it is a primary concern to develop an efficient computational method for systems with a high degree of freedom.
 Since most problems in condensed matter physics are difficult to solve analytically,
 a computational method plays an essential role in such situations.
 
 Markov chain Monte Carlo (MCMC), which enables us to sample spin configurations according to the Boltzmann distribution,
 has been successfully applied to various spin systems with a high degree of freedom \cite{newmanb}.
 Following the success, MCMC was combined with the renormalization group (RG) \cite{Wilson1,Wilson2}
 that is a key concept in understanding critical phenomena of phase transitions \cite{RGMC1,RGMC2,RGMC3}.
 The method, called Monte Carlo RG, numerically conducts Kadanoff's block-spin transformation in real space \cite{Kadanoff},
 and extracts essential information, such as critical exponents.
 It should also be noted that the inverse procedure of the block-spin transformation, called the inverse RG, was proposed by Ron, Swendsen, and Brandt \cite{IRG1,IRG2}.
 The inverse RG can generate larger size spin configurations by enlarging smaller ones without the critical slowing down.

 The rapid development of deep learning (DL) methods provides us with new paradigms to analyze condensed matter physics \cite{RevDLPhys1, RevDLPhys2, RevDLPhys3}.
 For example, the detection of a phase transition using DL has been done in various ways \cite{ptdl1,Carrasquilla2017,ptdl2,ptdl3,ptdl4,ptdl5,shiina2020,Tomita2020},
 and the relationship between RG and DL has been actively studied \cite{DLRG1,DLRG2,DLRG3,DLRG4,DLRG5,Li2018}. 
 Another application to spin models is to employ a DL model as the super-resolution method
 that is a technique enhancing the resolution of an image.
 Efthymiou {\it et al.} proposed using a convolutional neural network (CNN) to generate a larger size spin configuration from a smaller one \cite{Efthymiou2019}.
 It can be regarded as a way of inverse RG procedure.
 Although the method can explore a spin system on a large lattice without the critical slowing down,
 the problem is that it can be difficult for the CNN to generate proper samples above the critical temperature
 because the thermal noise is dominant in the region.
 To alleviate the problem, Shiina {\it et al.} proposed applying the super-resolution method to correlation configurations with the usage of improved estimator \cite{SRcorre}.

 Neural Network Renormalization Group (NNRG) \cite{Li2018} is a method to encode RG procedure into DL model.
 The method utilizes a flow-based generative model (or normalizing flow), one of the deep generative models. 
 As a flow-based generative model can exactly evaluate the likelihood, it can generate spin configurations from independent random noises without any data.
 Combining the model with MCMC simulation, one can obtain unbiased results. 
 It was shown that the convergence of the hybrid MCMC is faster than that of naive MCMC.
 However, it could be not easy when there exists a long-range correlation in the target system.
 This is because a flow-based generative model is not as expressive as the other generative models due to a strong restriction on its architecture.
 It should be noted that the idea is applied to a quantum system and lattice field theory \cite{xie2021abinitio, flowInlattice1, flowInlattice2} as well.

 in this study, we propose an application of a flow-based generative model to the super-resolution method.
 In addition to the noises, we feed smaller spin configurations to our flow-based generative model. 
 This helps the model to generate the larger configurations as the given smaller configurations have valuable information for the target spin system.
 We numerically show that our model can deal with a spin system on a larger lattice compared to NNRG.
 We also study transfer learning, where an optimized model is repeatedly used to explore further large systems step by step.
 This idea is inspired by the super-resolution with CNN \cite{Efthymiou2019}.
 The thermal averages of physical quantities obtained by our method have good agreements with the ones estimated by MCMC simulation.

 \section{Background}
 \label{sec:back}
  \subsection{Ising model}
  \label{sub:Ising}
  As a demonstration, we will apply our method to Ising model that is one of the fundamental models in condensed matter physics.
  For later convenience, we consider a two-dimensional $2L\times 2L$ square lattice with the linear system size $2L$.
  Each spin $s_i$ at the site $i$ belongs to $\mathbb{Z}_2$ which denotes a binary value,
  and a spin configuration is defined as $\bm{s}=\{s_1, s_2, \cdots, s_N\}$,
  where $N=2L\times 2L$ and $\bm{s}\in\mathbb{Z}_2^{N}$.
  The Boltzmann distribution of spin configuration is given by
  \begin{align}
   \label{eq:BD}
   \pi_s(\bm{s}) = \frac{1}{{\cal Z}}\exp{\left[\frac{1}{2}\bm{s}^t W \bm{s}\right]} \quad \text{with } \bm{s}\in\mathbb{Z}_2^{N}.
  \end{align}
  Here, the superscript $t$ indicates the transpose, and $W$ is a $N\times N$ symmetric matrix representing the dimensionless energy of nearest-neighbor interactions.
  The partition function is defined as ${\cal Z} = \sum_{\bm{s}\in \mathbb{Z}_2^{N}} \exp{\left[ \bm{s}^t W \bm{s}/2 \right]}$.

  \subsection{Continuous relaxation method}
  \label{sub:continuous}
   As most deep learning methods, including the flow-based generative model, have parameters that belong to a set of real numbers,
   it is not straightforward to deal with a discrete variable, for example, in Ising spin systems.
   In those cases, one can introduce the continuous relaxation method \cite{CR} that enables you to convert a discrete variable into a corresponding real variable.
   As a particular example of the method, we consider a two-dimensional Ising model of $N$($=2L\times 2L$) spins on a square lattice 
   in which the Boltzmann distribution is defined in the equation (\ref{eq:BD}).
   With an appropriate conditional probability $p(\bm{x}|\bm{s})$,
   one can introduce a real valued auxiliary vector $\bm{x}\in \mathbb{R}^N$
   in such a way that the interactions in the discrete variables are decoupled.
   Then, the marginal probability density function for the continuous variable is given by
   \begin{align}  
    \label{eq:pix}
    \piLL (\bm{x}) =& \sum_{\bm{s}} p(\bm{x}|\bm{s}) \pi_s(\bm{s}) \nonumber \\
    \propto& \exp{\left\{-\frac{1}{2} \bm{x}^T (W+\alpha I)^{-1}\bm{x} \right\}} \prod_{i}^N\cosh{(x_i)} \\
    & \text{with } \bm{x}\in \mathbb{R}^N, \nonumber
   \end{align}
   where $I$ is $N\times N$ identical matrix and $\alpha$ is a real value for which $W+\alpha I$ is a positive definite.
   Using Bayes' theorem, we also obtain
   \begin{align}  
    \label{eq:psx}
    p(\bm{s}|\bm{x}) = \prod_i^N (1+e^{-2s_ix_i})^{-1}.
   \end{align}
   Note that $p(\bm{s}|\bm{x})$ is factorized into the product of elements. 
   Therefore, one can estimate some thermal average of physical quantities in regard to $\bm{s}$ by using $\bm{x}$.
   In the end, instead of dealing with $\bm{s}$, 
   we can consider $\bm{x}$ as a target variable with the application of deep learning methods.
   Hereafter, we assume $\bm{x}$ to be spin configuration and $\piLL (\bm{x})$ to be Boltzmann distribution unless otherwise stated.
   
  \subsection{Flow-based generative model}
  \label{sub:Flow}
   The flow-based generative model, initially proposed by Dinh {\it et al.} \cite{Dinh2015}, is one of the deep generative models that makes use of the invertible property.
   This property leads to tractable inference and log-likelihood.
   However, instead, it has strong restrictions to its architecture,
   and the function space that the model can explore is limited, compared to the other generative models, 
   such as Variational Autoencoders, Generative Adversarial Network, and Autoregressive Model.

   We define latent variable $\bm{z}\in \mathbb{R}^{N}$ and data point $\bm{x}\in \mathbb{R}^{N}$.
   Each of them follows $p$ and $p_{data}$, which are denoted by $\bm{z}\sim p$ and $\bm{x}\sim p_{data}$, respectively.
   The distribution $p$ is an arbitrary latent distribution, and $p_{data}$ is an unknown data distribution.
   To describe the flow-based generative model, 
   we consider a bijective function $f_{\theta}:\mathbb{R}^{N}\rightarrow \mathbb{R}^{N}$ parameterized by $\bm{\theta}$ and its inverse function $f_{\theta}^{-1}$.
   The forward and backward equations of the flow-based generative model are given respectively by
   \begin{align}
    \left\{
    \begin{aligned}
     \bm{x} &= f_{\theta}(\bm{z})\\
     \bm{z} &= f_{\theta}^{-1}(\bm{x}).
    \end{aligned}
    \right.
   \end{align}
   The change of variable formula leads to the model distribution $q_{\theta}$;
   \begin{align}
    \label{eq:model}
    q_{\theta}(\bm{x})
    &= p(f_{\theta}^{-1}(\bm{x})) \left|\det{ \left(\frac{\partial f_{\theta}^{-1}(\bm{x})}{\partial \bm{x}}\right)}\right|,
   \end{align}
   where $\frac{\partial f_{\theta}^{-1}(\bm{x})}{\partial \bm{x}}$ is a Jacobian matrix of $f_{\theta}^{-1}$ at $\bm{x}$.
   The log-likelihood for a data point is written as
   \begin{align}
    \ln{q_{\theta}(\bm{x})} = \ln{p(f_{\theta}^{-1}(\bm{x}))} + \ln{\left|\det{ \left(\frac{\partial f_{\theta}^{-1}(\bm{x})}{\partial \bm{x}}\right)}\right|}.
   \end{align}
   With a data set and these equations, we can explicitly estimate $q_{\theta}$ and its log-likelihood.
   However, the determinant of Jacobian is computationally intractable, so Dinh {\it et al.} designed the architecture such that the Jacobian will be a triangular matrix.
   The extension of this original idea has been studied by changing the detailed designs of the architecture \cite{Kingma2018,Dinh2019}.
   Note that, in the rest of this paper, 
   the symbol $\bm{x}$ defined in this section will be the same $\bm{x}$ of the continuous spin configuration introduced in the section \ref{sub:continuous},
   and $p_{data}$ will correspond to its Boltzmann distribution $\piLL$.
   For the latent variable $\bm{z}$, Li and Wang \cite{Li2018} use an isotropic gaussian as explained in the next section.
   While $\bm{z}$ in the context of this study will be a concatenation of $\myxL$ and $\myg$ which we will define in the section \ref{sub:optimization} later.
  
  \subsection{Neural Network Renormalization Group (NNRG)}
  \label{sub:NNRG}
   Before going into the detail of our work, we introduce the Neural Network Renormalization Group (NNRG) proposed by Li and Wang 
   that is an application of a flow-based generative model to physics problems \cite{Li2018}. Our method based on the NNRG will be presented in the next section.

   When considering a physics system, we know that the Boltzmann distribution governs a target physical variable.
   It enables us to employ kullback leibler (KL) divergence between the model distribution $q_{\theta}$ and the Boltzmann distribution $\piLL$
   as the objective function rather than log-likelihood.
   Therefore, the NNRG does not require any data, and the unsupervised learning in a flow-based generative model turns into an optimization problem.
   As is the case in Li and Wang \cite{Li2018},
   let the continuous ising spin configuration $\bm{x}\in \mathbb{R}^{N}$ be the target physical variables,
   and the gaussian noise $\bm{\gamma}\in \mathbb{R}^{N}$ be the latent variables
   where $\bm{\gamma}$ follows the multivariate gaussian with the diagonal covariance matrix (the isotropic gaussian) denoted by $p_g$.
   Given the equation (\ref{eq:model}) with a property of Jacobian; $|J_{f}|=|J_{f^{-1}}|^{-1}$,
   the KL divergence for the optimization can be written as 
   \begin{align}
    \label{eq:L}
    {\cal L}_{\theta}
    &= E_{\bm{\gamma}\sim p} \left[\ln{p_g(\bm{\gamma})} - \ln{\left|\det{ \left(\frac{\partial f_{\theta}(\bm{\gamma})}{\partial \bm{\gamma}}\right) }\right|}
    - \ln{\tilde{\pi}_{x}(f_{\theta}(\bm{\gamma}))}\right].
   \end{align}
   Here, $\tilde{\pi}_{x}(\bm{x})$ is the unnormalized Boltzmann distribution, and the term of the partition function is omitted because it does not depend on the parameters $\bm{\theta}$.
   The minimization of the objective function with respect to $\bm{\theta}$ leads to the model distribution 
   $q_{\theta}$ that approximates the target Boltzmann distribution $\tilde{\pi}_{x}$.

   The numerical optimization with the importance sampling for the equation (\ref{eq:L}) can be insufficient, so Li and Wang combined the optimized model with MCMC simulation.
   To be precise, the partition function is expressed as 
   \begin{align}
    \label{eq:Z}
    {\cal Z}_x 
    &= \int \tilde{\pi}_{x}(\bm{x}) d\bm{x} \nonumber \\
    &= \int p_g(\bm{\gamma}) \frac{ \tilde{\pi}_{x}(f_{\theta}(\bm{\gamma})) }{ q_{\theta}(f_{\theta}(\bm{\gamma})) } d\bm{\gamma},
   \end{align}
   where the variable $\bm{x}$ is changed to $\bm{\gamma}$, and the equation (\ref{eq:model}) is used.
   Since the partition function is expressed only by the latent variable $\bm{\gamma}$,
   the effective energy function in the latent space can be written as
   $E_z(\bm{\gamma}) = -\ln{p_g(\bm{\gamma})} -\ln{ \tilde{\pi}_{x}(f_{\theta}(\bm{\gamma})) } + \ln{ q_{\theta}(f_{\theta}(\bm{\gamma})) }$.
   This equation enables us to perform the Hamiltonian Monte Carlo (HMC) sampling \cite{HMC} in the latent space.
   The energy represented by the latent variable can be simpler than the original one if $\tilde{\pi}_{x}\simeq q_{\theta}$,
   which makes HMC in latent space faster to converge.

 \section{Method}
 \label{sec:Method}
 \begin{figure}[t]
  \centering
  \includegraphics[width=1.0\linewidth]{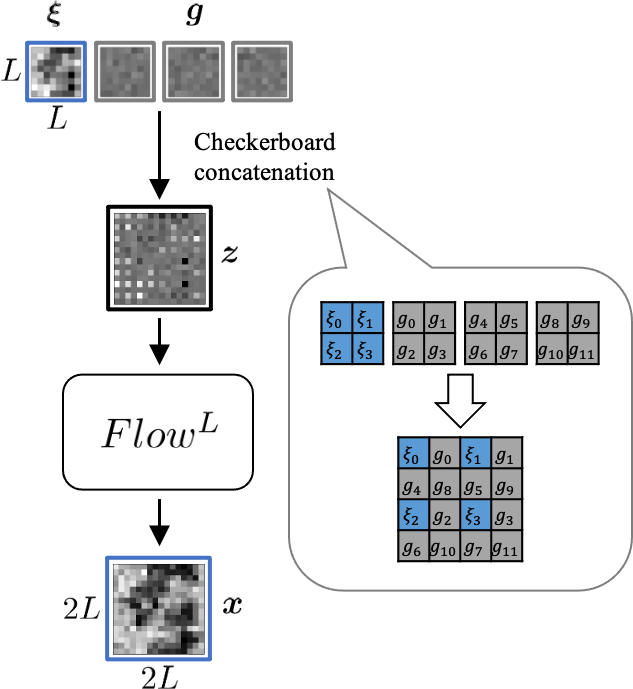}
  \caption{\label{fig:flow} The illustration of our main idea.
  The concatenation of an $L\times L$ spin configuration $\myxL$ (blue frame) and Gaussian noises $\myg$ (gray frames)
  is passed into the flow-based generative model (denoted by $Flow^{L}$).
  Then, it generates $2L\times 2L$ spin configuration $\myxLL$ (blue frame).
  For the concatenation, we use the checkerboard concatenation whose example with $L=2$ is depicted in the box.
  Here, $\xi_i$ and $g_i$ are the elements of $\myxL$ and $\myg$, respectively.}
 \end{figure}  
 The NNRG succeeded in generating two-dimensional Ising spin configurations from simple Gaussian noises.
 However, since a flow-based generative model is not as expressive as the other generative models are, the optimization will be challenging when dealing with more complex data space with large lattice sizes or when the target spin system has more complicated interactions.
 
 In this paper, we propose using the NNRG as a super-resolution method. 
 Instead of giving only Gaussian noises to the flow-based generative model, we give spin configurations of a smaller size $L\times L$  with some Gaussian noises.
 Then, the model generates spin configurations of an increased size $2L\times 2L$.
 The information contained in the input smaller configurations helps the model learn the distribution of the larger configurations.
 The illustration of this idea is depicted in Fig. \ref{fig:flow}.

  \subsection{Optimization}
  \label{sub:optimization}
  To generate an increased size configuration $\myxLL \in \mathbb{R}^{2L\times 2L}$,
  we input a latent variable $\myz \in \mathbb{R}^{2L\times 2L}$ that is the concatenation of an $L\times L$ spin configuration $\myxL \in \mathbb{R}^{L\times L}$
  and an isotropic gaussian noise $\myg \in \mathbb{R}^{3(L\times L)}$.
  For the concatenation, we employ the checkerboard concatenation where the noises are inserted between the elements of $\myxL$.
  The illustration of this procedure for $L=2$ is depicted in the box on Fig.~\ref{fig:flow}.
  This way of concatenation is inspired by Kadanoff's block-spin transformation \cite{Kadanoff}.
  Note that the latent variable $\bm{z}$ does not follow $p_g$ any more.
  Since $\myz$ is the concatenation of independent variables $\myxL$ and $\myg$, its distribution $p_z$ is given by
  \begin{equation}
   \label{eq:p_z}
   p_z(\myz) = \piL(\myxL) p_g(\myg).
  \end{equation}
  Here, $\piL$ is the Boltzmann distribution for $L\times L$ lattice at temperature $T_{\xi}$.
  The forward and backward equations of our flow-based generative model, parametrized by $\bm{\theta}$, are given by
  \begin{align}
   \left\{
   \begin{aligned}
    \myxLL &= f_{\theta}(\myz)\\
    \myz &= f_{\theta}^{-1}(\myxLL)
   \end{aligned}
   \right.
  \end{align}
  By using the equation (\ref{eq:p_z}), the model distribution of $\myxLL$ will be of the form,
  \begin{align}
   \label{eq:q_theta}
   q_{\theta}(\myxLL) &= p_z(\myz) \left|\det{ \left(\frac{\partial f_{\theta}(\myz)}{\partial \myz}\right) }\right|^{-1}.
  \end{align}
  
  With $\myz\sim p_z$, our model will be optimized so that $q_{\theta}$ approximates $\piLL$ 
  which is the Boltzmann distribution for $2L\times 2L$ lattice at target temperature $T_{x}$.
  As with the equation (\ref{eq:L}), the objective function is defined by
  \begin{widetext}
   \begin{align}
    \label{eq:L2}
    {\cal L}_{\theta}
    &= \int p_{z}(\myz) \left[\ln{p_{z}(\myz)} - \ln{\left|\det{ \left(\frac{\partial f_{\theta}(\myz)}{\partial \myz}\right) }\right|}
    - \ln{\tpiLL(f_{\theta}(\myz))} \right] d\myz,
   \end{align}
  \end{widetext}
  where $\tpiLL$ is the unnormalized Boltzmann distribution. 
  The data set $\{\bm{\xi}^{(d)}\}_{d=1}^{D}$ of $D$'s samples is provided by the naive HMC and is used to approximate the integral of equation (\ref{eq:L2}) during a numerical optimization.
  Since we do not know the partition function of $\piL$,
  it is replaced by the unnormalized one $\tilde{\pi}_{\xi}$.
  This will not cause any problem in the optimization 
  because the partition function does not depend on $\bm{\theta}$ and is just a constant.
  
  To help the model quickly learn the super-resolution mapping, we add to the equation (\ref{eq:L2}) a regularization term,
  $\gamma \left|\langle m\rangle_{in} - \langle m\rangle_{ds} \right|$, 
  where $\gamma$ is a hyperparameter, and the magnetization per spin is defined as $m=\left(L\times L\right)^{-1}\sum_{i}^{L\times L}s_i$.
  The angular bracket $\langle\cdot\rangle_{in}$ represents the sample average over the input configurations $\{\bm{\xi}^{(d)}\}_{d=1}^{D}$.
  Although $m$ depends on the discrete spin variables, $\langle m\rangle_{in}$ can be estimated from the continuous spin variables
  because the conditional probability of $\bm{s}$ given $\bm{\xi}$ is factorized into the product of elements as one can see from equation (\ref{eq:psx}).
  While the angular bracket $\langle\cdot\rangle_{ds}$ is the sample average over the output configurations $\{\bm{x}^{(d)}\}_{d=1}^{D}$ with the consideration of down-sampling.
  It means that we estimate $\langle m\rangle_{ds}$ over the probability of down-sampled discrete spin variables associated with $\{\bm{x}^{(d)}\}_{d=1}^{D}$.
  This can be done because the down sampling, for which we use the majority rule on $2\times 2$ window, is a local operation,
  and one can systematically list up all possibilities of down-sampling processes on a $2\times 2$ window.
  The regularization term forces the magnetization of down-sampled output configurations to be close to that of input configurations.
  It helps the model encode the super-resolution mapping between $\bm{\xi}$ and $\bm{x}$.
  In our numerical experiments, we use $\gamma=200$
  and make it zero after a few steps of optimization to get the pure optimization of equation (\ref{eq:L2}) in the end.

  The detailed architecture of our flow-based generative model is given in Appendix \ref{app:arch}.

  \subsection{Approximate temperature rescaling}
  \label{sub:atr}
  \begin{figure}[t]
   \centering
   \includegraphics[width=0.90\linewidth]{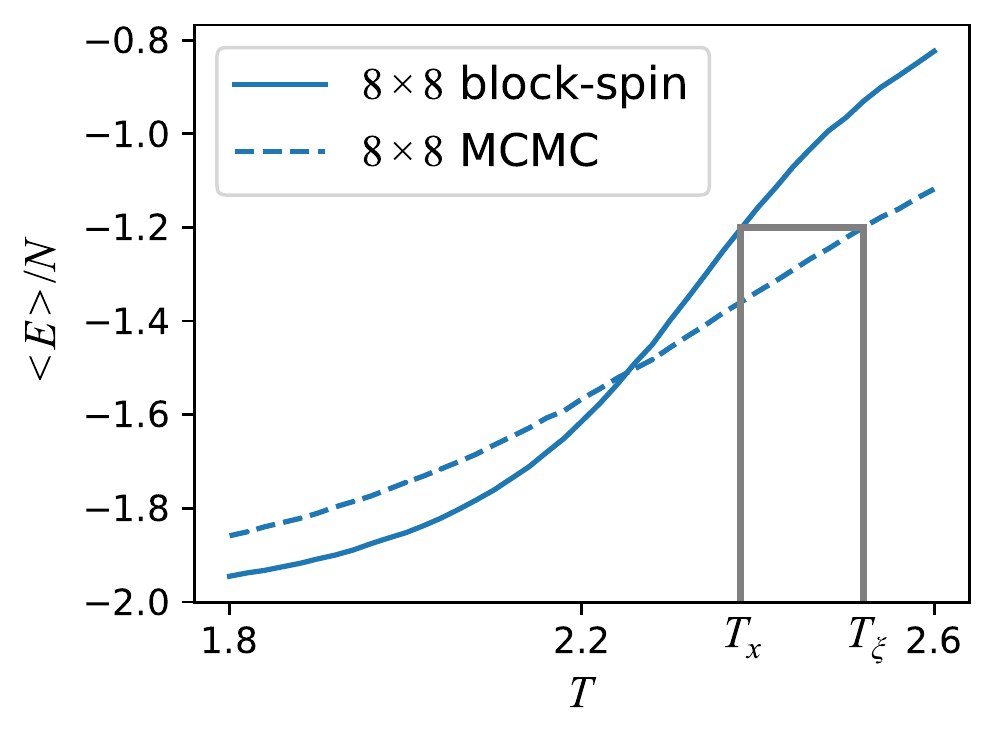}
   \caption{\label{fig:ar} This figure illustrates the way of the approximate temperature rescaling in the case $L=16$.
   The solid line is the thermal average of energy density for the $L/2\times L/2$ block-spin configurations obtained from $L\times L$ spin configurations.
   The dotted line is the energy density for the $L/2\times L/2$ spin configurations.
   Given the target temperature $T_{x}$, look at the energy of the block-spin configurations 
   and find the input temperature $T_{\xi}$ where the energy of $L/2\times L/2$ configurations have the same value.}   
  \end{figure}
  In order to support the model to learn the Boltzmann distribution on $2L\times 2L$ lattice at temperature $T_{x}$,
  we need to carefully choose $T_{\xi}$ that is the temperature for input $L\times L$ configurations.
  When a configuration at temperature $T$ is transformed by the Kadanoff's block-spin transformation,
  we will obtain a block-spin configuration at effective temperature $\tilde{T}=h(T)$ with unknown function $h$.
  As our flow-based generative model is supposed to realize the super-resolution,
  the temperature $T_{\xi}=h^{-1}(T_{x})$ would be a good approximation for the input temperature, where $h^{-1}$ is the inverse function of $h$.
  Thus, we numerically establish $h$ by MCMC samples and apply it to infer $T_{\xi}$.
  This method was proposed in \cite{Efthymiou2019} and called the approximate temperature rescaling.

  In our scheme, the $L\times L$ configurations are sampled from MCMC simulation over some temperature range,
  and the $2\times 2$ majority rule is applied to them to obtain the $L/2\times L/2$ block-spin configurations.
  Then, comparing the energy density of the block-spin configurations with that of $L/2\times L/2$ spin configurations,
  we can numerically establish the transformation $h$ and its inverse one $h^{-1}$.
  Once the transformations are established on $L\times L \rightarrow L/2\times L/2$,
  using the same transformation, we approximately infer $T_{\xi}$ that is temperature for $L\times L$ configuration
  appropriate for the generating $2L\times 2L$ configurations (Fig.~\ref{fig:ar}).

  \subsection{Efficient HMC with a flow-based generative model}
  \label{sub:HMC-Flow}
  As suggested in \cite{Li2018}, one can perform HMC in a simpler latent space with the application of a flow-based generative model.
  In our scheme, we can reduce the complexity of the target variable $\myxLL$, where the correlation length is $2L$,
  to that of the latent variable $\myz$ with the shorter correlation length $L$.
  Similar to the equation (\ref{eq:Z}), the partition function for the Boltzmann distribution $\piLL$ can be written as 
  \begin{align}
   \label{eq:Z2}
   {\cal Z}
   &= \iint \piL(\myxL) p_g(\myg) \frac{ \tpiLL(f_{\theta}(\myz)) }{ q_{\theta}(f_{\theta}(\myz)) } d\myxL d\myg.
  \end{align}
  Note that $\myz$ is the concatenation of $\myxL$ and $\myg$.
  Thus, the effective energy in the latent space will be
  $E_z(\myz) = -\ln{\piL(\myxL)} - \ln{p_g(\myg)} -\ln{ \tpiLL(f_{\theta}(\myz)) } + \ln{ q_{\theta}(f_{\theta}(\myz)) }.$
  Using this effective energy, we can perform an efficient HMC in the latent space where the convergence is faster than that for the original target space.

  \subsection{Transfer learning and extrapolation}
  \label{sub:ext}
  \begin{figure}
   \includegraphics[width=0.80\linewidth]{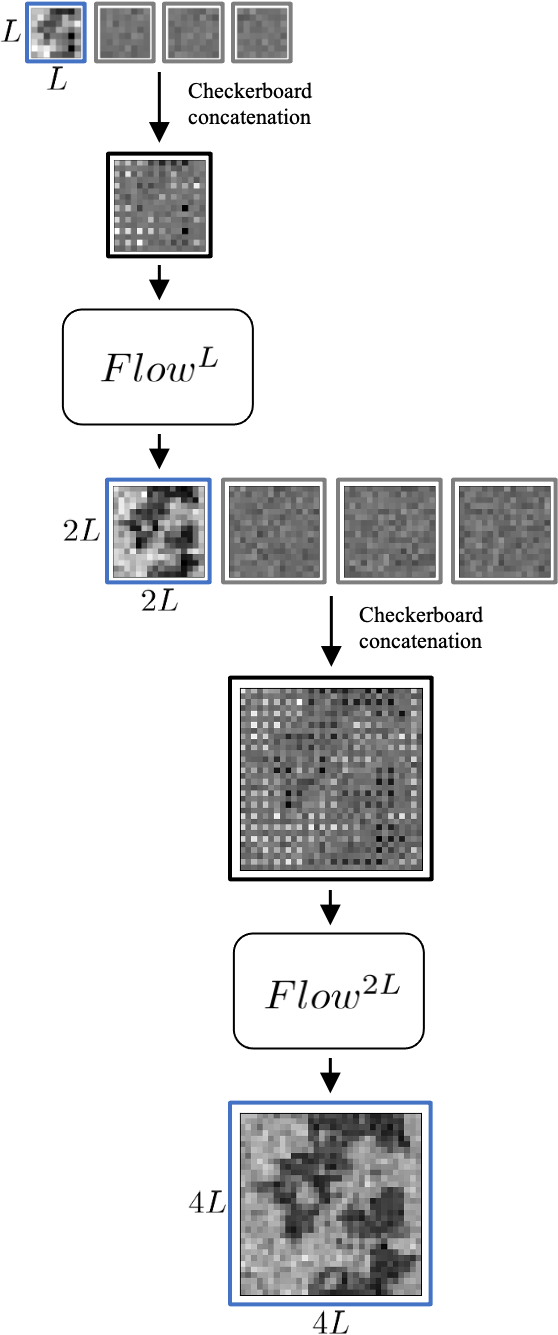}
   \caption{\label{fig:trans} In the transfer learning, we optimize $Flow^{L}$ and then use its parameters as the initial parameters for $Flow^{2L}$.
   $Flow^{2L}$ is optimized by using $\{\myxLL^{(d)}\}_{d=1}^{D}$ generated by $Flow^{L}$.}
  \end{figure}
  We define $Flow^{L}$ as the model that generates $2L\times 2L$ configurations from $L\times L$ configurations.
  The optimized parameters in $Flow^{L}$ can be used as the initial parameters for $Flow^{2L}$ which generates $4L\times 4L$ configurations.
  Here, the data set $\{\myxLL_i\}_{d=1}^{D}$ to optimize $Flow^{2L}$ can be provided by $Flow^{L}$.
  Although the sizes of configurations that $Flow^{L}$ and $Flow^{2L}$ deal with are different, 
  the architecture of our model can be adapted to any size of configuration (see Appendix \ref{app:arch} for the detail of architecture).
  This feature enables us to reuse the parameters,
  and the procedure is generally called transfer learning \cite{transfer}. 
  Applying the transfer learning step by step, we can explore larger size systems. The way of our transfer learning scheme is illustrated in Fig.~\ref{fig:trans}.

 \section{Numerical results}
 \label{sec:results}
 Here we show the results of numerical experiments for a two-dimensional Ising model with $L=16$.
 After preparing the data set of $L\times L$ configurations with the naive HMC simulation, 
 we feed it to $Flow^{L}$ and optimize its model parameters $\bm{\theta}$ by minimizing the objective function defined in the equation (\ref{eq:L2}).
 The optimized model is used to perform the efficient HMC (explained in the section \ref{sub:HMC-Flow}) in order to obtain sophisticated results.

  \subsection{Convergence of the efficient HMC}
\begin{figure}[t]
   \includegraphics[width=0.9\linewidth]{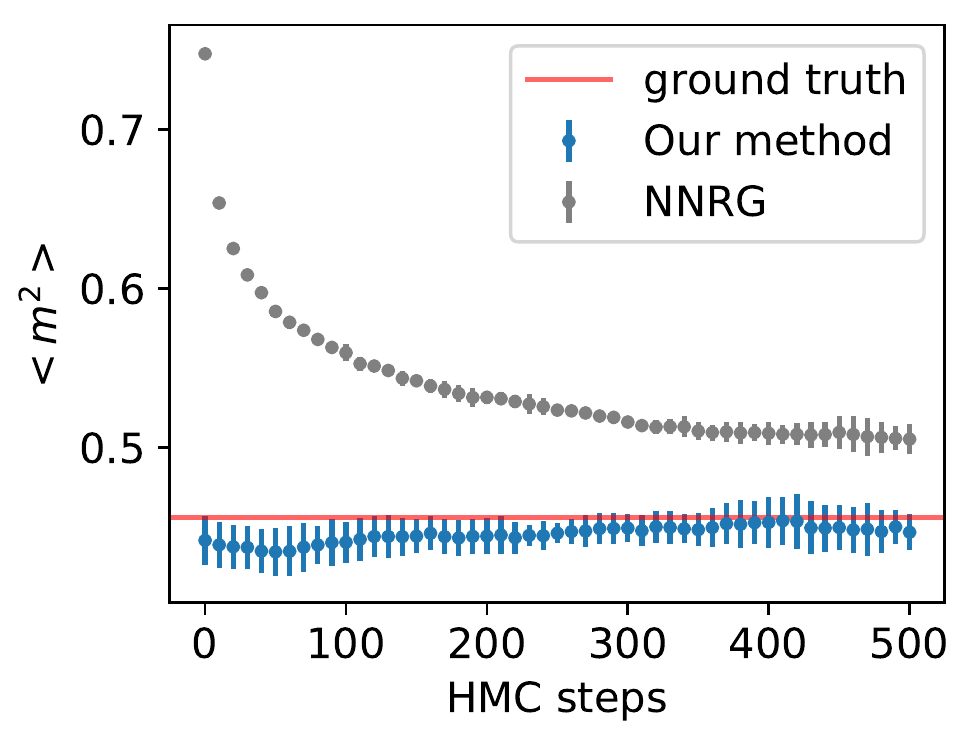}
   \caption{\label{fig:hmc} The structure factors on a $32\times 32$ lattice at the critical temperature ($2.269...$) over the HMC steps. 
   The result of our method (blue dots) converges to the ground truth (red line) faster than that of NNRG (gray dots).
   The ground truth is obtained by MCMC simulations with Swendsen-Wang cluster update \cite{SW} in the discrete spin space.}     
  \end{figure}

  We compare our method with NNRG \cite{Li2018} by looking at the convergence of HMC simulation with the optimized models.
  Fig.~\ref{fig:hmc} shows the structure factors for each step of the efficient HMC simulation on a $32\times 32$ lattice.
  NNRG generates $32\times 32$ configurations directly from the gaussian noises,
  while our method utilizes the information of $16\times 16$ configurations.
  We can see that the result of our method converges to the ground truth much faster than that of NNRG.
  This result indicates that our method could be a better choice when the original input spin system $L\times L$ at temperature $T_{\xi}$ does not suffer from the critical slowing down.
  Note that Li and Wang \cite{Li2018} made the comparison of the convergences between naive HMC and NNRG,
  and showed that NNRG has better convergence.

  Another advantage of our method is that the total number of parameters in our model is about $40\%$ less than that used in NNRG.
  In our numerical experiments, NNRG uses 63200 parameters while our model includes 38400 parameters.

  \subsection{Results of extrapolation}
\begin{figure}[t]
   \includegraphics[width=0.9\linewidth]{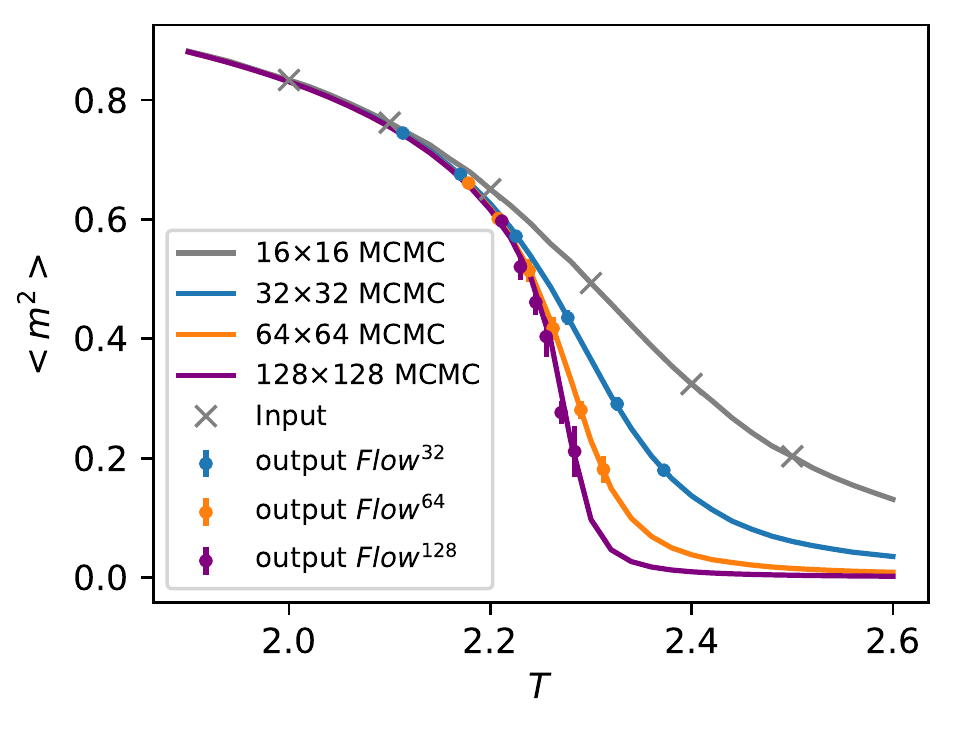}
   \caption{\label{fig:ext} The structure factors obtained by the efficient HMC simulation with the optimized flow-based generative models.
   The solid lines are obtained from MCMC simulations with Swendsen-Wang cluster update \cite{SW} in the discrete spin space.
   Giving $16\times 16$ configurations to $Flow^{16}$ at each temperature (depicted by gray dots), we obtain the structure factors on $32\times 32$ lattice (blue crosses).
   The structure factors on $64\times 64$ lattice (orange crosses) are then obtained from $Flow^{32}$ using the transfer learning and extrapolation introduced in the section \ref{sub:ext}.}
  \end{figure}

  After optimizing $Flow^{L}$, we use its optimized parameters $\bm{\theta}$ as the initial parameters for $Flow^{2L}$
  and optimize them using the data set $\{\bm{\myxLL}^{(d)}\}_{d=1}^{D}$ where $\myxLL^{(d)}\sim q_{\theta}$.
  Combining both $Flow^{L}$ and $Flow^{2L}$ with HMC simulation, 
  we estimate the structure factor, $\langle m^2\rangle=\langle \sum_{i,j}s_is_j\rangle / (4L\times 4L)^2$, on $4L\times 4L$ lattice.
  We iterate the extrapolation procedure step by step and obtain the samples on $128\times 128$, starting from $16\times 16$.
  The structure factors obtained by our method are shown in Fig.~\ref{fig:ext}.
  We can see that these have good agreement with the results of MCMC simulation in the discrete spin space.
  These results are distinctive from the ones in the super-resolution of CNN \cite{Efthymiou2019}.
  Our results show a good agreement even at high temperature
  because the method can estimate the model distribution, and it enables us to obtain unbiased results with the efficient HMC.

 \section{Conclusion}
 \label{sec:conc}
 We proposed a new paradigm of super-resolution for spin configurations with the application of a flow-based generative model.
 Starting from a data set of small-size spin configurations, our model provides larger-size configurations in a faster computation.
 As our method exactly evaluates the distribution of the generated spin configurations,
 it can be combined with HMC simulation, which efficiently converges to the Boltzmann distribution.
 In the resulting efficient HMC, our flow-based generative model reduces the correlation length on the target spin variable 
 to the shorter one on the original input spin variable.

 We numerically showed that our method could efficiently simulate the Ising model on a two-dimensional square lattice.
 We also implemented ``transfer learning'' to explore the larger-size systems.
 Using a data set composed of $16\times 16$ spin configurations, 
 our model provides the 8 times larger configurations ($128\times 128$).
 The thermal average of physical quantities obtained by the method showed a good agreement with the one obtained from the MCMC simulations.

 For future research,
 a spin system with multi-component spin variables could be an interesting target for the application of our method.
 For instance, the $q$-state clock model (XY model) is known to have the Berezinskii-Kosterlitz-Thouless (BKT) transition \cite{BKT1,BKT2,BKT3,BKT4} 
 where the quasi long-range order exists.
 With the application to this model, we can see whether our method is valid for a phase transition without symmetry breaking.
 These spin systems were also studied in the super-resolution method with CNN using correlation configurations \cite{SRcorre}.


%



\begin{acknowledgments}
 We would like to thank Liu Wei and Sojeong Park for proof reading of our paper and giving valuable comments. 
 This work was supported by a Research Fellowships of Japan Society for the Promotion of Science for Young Scientists, Grant Number20J12472.
 K. S. is also grateful to the A*STAR (Agency for Science, Technology and Research) Research AttachmentProgramme of Singapore for financial support.
\end{acknowledgments}

\appendix
 \section{Architecture}
 \label{app:arch}
 \begin{figure}[t]
  \includegraphics[width=\linewidth]{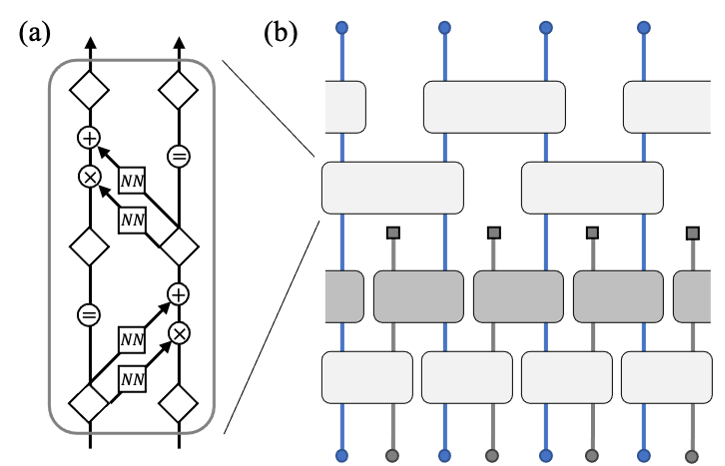}
  \centering
  \caption{\label{fig:arch} (a) The illustration of a bijector block with two coupling layers.
  The square with $NN$ denotes a fully-connected neural network, and a coupling layer consists of two fully-connected neural networks.
  The open diamonds are nodes, and the circles represent arithmetic operations.
  (b) The simplified one-dimensional illustration of our flow-based generative model based on NNRG \cite{Li2018}.
  The boxes are bijector blocks. Among these, dark gray boxes are disentanglers that make variables disentangled.}
 \end{figure}
 For our flow-based generative model, we hired the architecture of Neural Network Renormalization Group (NNRG) developed in \cite{Li2018}.
 To simplify the discussion, we consider the architecture in one dimension because the two-dimensional case, which is used in this study, is essentially the same.

 NNRG architecture is composed of some bijector blocks, each of which takes two variables as inputs and gives two outputs (it is four instead of two in a two-dimensional case).
 Each bijector, as initially proposed in \cite{Dinh2015}, is bijective and consists of coupling layers.
 FIG. \ref{fig:arch}(a) illustrates an example architecture of a bijector block with two coupling layers, each of which includes two fully-connected neural networks.
 Stacking such bijector blocks into layers, one can realize the invertible transformation between two variable spaces.
 FIG. \ref{fig:arch}(b) is a simple example with 4 layers.
 The straight blue lines go through all layers while gray lines through only two layers.
 This is an essential part of NNRG architecture where the variables represented by gray squares on the top of gray lines are disentangled.
 The bijector blocks colored with dark gray do it and are called disentangler in \cite{Li2018}.
 In our case, the variables represented by gray squares correspond to the gaussian noises.

 For the straightforward extension to the two-dimensional case, one can refer to \cite{Li2018}.
 In our numerical experiments, we use 16 layers in total where the layer for disentangling is at the 8th layer from the bottom.
 A bijector includes 10 coupling layers, 
 and the fully-connected neural networks in a coupling layer consist of 3 layers with 4 inputs nodes, 6 hidden nodes, and 4 output nodes.
 We chose Exponential Linear Unit (ELU) as the activation function for the input and hidden layers.
 The bijectors in the same layer share weights. This is suitable for the target spin systems because of translational invariances.
 In addition, it enables us to apply the architecture to spin systems of any linear size.

 \section{Formulations for extrapolation}
 \label{app:HMC-Flow}
 In the section \ref{sub:ext}, we considered the transfer learning and extrapolation. 
 Here, we summarize the formulations for $Flow^{2L}$ that generates a $4L\times 4L$ spin configuration $\bm{x}'\in \mathbb{R}^{4L\times 4L}$.
 The input to the $Flow^{2L}$, denoted by $\bm{z}'\in \mathbb{R}^{4L\times 4L}$, 
 is the concatenation of $\bm{x}\in \mathbb{R}^{2L\times 2L}$ and the gaussian noise $\bm{g}' \in \mathbb{R}^{3(2L\times 2L)}$.
 As $\bm{x}$ and $\bm{g}'$ are independent of each other, $\bm{z}'$ follows
 \begin{align}
  p_{z'}(\bm{z}') = \pi_{x}(\bm{x}) p_g(\bm{g}').
 \end{align}
 Note that the dash symbols are not the symbol of differentiation.
 Then, the forward and backward equations of $Flow^{2L}$ are defined respectively by
 \begin{align}
  \left\{
  \begin{aligned}
   \bm{x}' &= h_{\nu}(\bm{z}')\\
   \bm{z}' &= h_{\nu}^{-1}(\bm{x}'),
  \end{aligned}
  \right.
 \end{align}
 where $h_{\nu}$ is the model function of $Flow^{2L}$ parameterized by $\nu$.
 Similar to the equation (\ref{eq:q_theta}), the model distribution of $\bm{x}'$ is given by
 \begin{align}
  \label{eq:r_nu}
  r_{\nu}(\bm{x}') &= p_{z'}(\bm{z}') \left|\det{ \left(\frac{\partial h_{\nu}(\bm{z}')}{\partial \bm{z}'}\right) }\right|^{-1}.
 \end{align}
 With the above equations, the partition function of Ising model on $4L\times 4L$ lattice can be written as 
  \begin{align}
   {\cal Z}_{x'}
   &= \int \tilde{\pi}_{x'}(\bm{x}') d\bm{x}' \nonumber \\
   &= \int p_{z'}(\bm{z}') \frac{ \tilde{\pi}_{x'}(h_{\nu}(\bm{z}')) }{ r_{\nu}(h_{\nu}(\bm{z}')) } d\bm{z}' \nonumber \\
   &= \iint \tilde{\pi}_{x}(\myxLL)p_{g}(\bm{g}') \frac{ \tilde{\pi}_{x'}(h_{\nu}(\bm{z}')) }{ r_{\nu}(h_{\nu}(\bm{z}')) } d\myxLL d\bm{g}' \nonumber \\
   &= \iiint \tilde{\pi}_{\xi}(\myxL) p_{g}(\bm{g}) p_{g}(\bm{g}') \frac{ \tilde{\pi}_{x'}(h_{\nu}(\bm{z}')) }{ r_{\nu}(h_{\nu}(\bm{z}')) } 
   \frac{ \tilde{\pi}_{x}(q_{\theta}(\bm{z})) }{ q_{\theta}(f_{\theta}(\bm{z})) }  d\myxL d\bm{g}d\bm{g}',
  \end{align}
 where $\tilde{\pi}_{x'}$ is the unnormalized Boltzmann distribution on $4L\times 4L$ lattice.
 Using the internal equation of the integral, we can perform an efficient HMC in the latent space to sample $\bm{x}'$.
 When both of the model distributions $f_{\theta}$ and $r_{\nu}$ approximate the Boltzmann distributions well, the convergence of the efficient HMC will be faster than that in the original space.

\bibliography{main}
\bibliographystyle{apsrev4-2}

\end{document}